# Ultra-broadband extreme-ultraviolet lensless imaging of extended complex structures

Stefan Witte[*,] Vasco T. Tenner, Daniël W. E. Noom, Kjeld S. E. Eikema

LaserLaB Amsterdam, VU University, De Boelelaan 1081, 1081 HV Amsterdam, The Netherlands

[*]Correspondence to: s.m.witte@vu.nl

**Lensless imaging**[1–3] **is an elegant approach to high-resolution microscopy, which is rapidly gaining popularity in applications where imaging optics are problematic. However, current lensless imaging methods require objects to be placed within a well-defined support structure**[4,5]**, while the light source needs to have a narrow, stable, and accurately known spectrum**[6]**.**

**Here we introduce a general approach to lensless imaging without spectral bandwidth limitations or sample requirements. We use two time-delayed coherent light pulses, and show that scanning the pulse-to-pulse time delay allows the reconstruction of diffraction-limited images for all spectral components in the pulse. Moreover, an iterative phase retrieval algorithm is introduced, which uses these spectrally resolved Fresnel diffraction patterns to obtain high-resolution images of complex extended objects without any support requirements.**

**We demonstrate this two-pulse imaging method with octave-spanning visible light sources (in both transmission**[2] **and reflection**[7] **geometries), and with broadband extreme-ultraviolet radiation from a high-harmonic source. This demonstrates that our approach enables effective use of low-flux ultra-broadband sources, such as table-top soft-X-ray systems**[8–10]**, for high-resolution imaging.**

The central issue in lensless imaging is the retrieval of phase information from a recorded diffraction pattern. Solutions to this problem have been found both through numerical and optical means. In coherent diffractive imaging, the missing phase information is numerically reconstructed by iterative algorithms[11], which has been shown to have a unique solution as long as the oversampling condition is satisfied[12]. Holographic methods, which directly record the phase through interference with a separate reference wave, have also been developed in a lensless imaging context[3]. In general, the use of monochromatic radiation has always been a major requirement for any diffractive imaging experiment, since the angle at which light diffracts from any structure depends on its wavelength. A diffraction pattern generated by a polychromatic source will therefore consist of a superposition of diffraction patterns of all the individual spectral components[13]. Depending on the spectral bandwidth of the source, this effect limits the achievable resolution and, for more broadband sources, will inhibit image reconstruction entirely.

There is a growing need for lensless imaging methods that operate with broadband radiation, to make much more efficient use of state-of-the-art light sources such as 3[rd] generation synchrotrons and table-top high-harmonic generation (HHG) sources for imaging. Especially HHG-based sources hold promise for compact and cost-effective ultrahigh-resolution microscopy, and recent breakthroughs have led to the generation of a significant photon flux in the biologically important water-window spectral range[9]. However, the need to spectrally filter these intrinsically ultra-broadband HHG sources leads to a reduction in flux of 3-4 orders of magnitude, which strongly hampers the use of these table-top sources for imaging applications. By incorporating knowledge of the source spectrum into the reconstruction algorithm, discrete spectra[13] and finite bandwidths[6] can be handled. But such calculations require a stable, accurately calibrated source spectrum, and therefore remains limited to relatively narrow spectral bandwidths ($\Delta\lambda/\lambda \sim 0.028$ has been shown[6]).

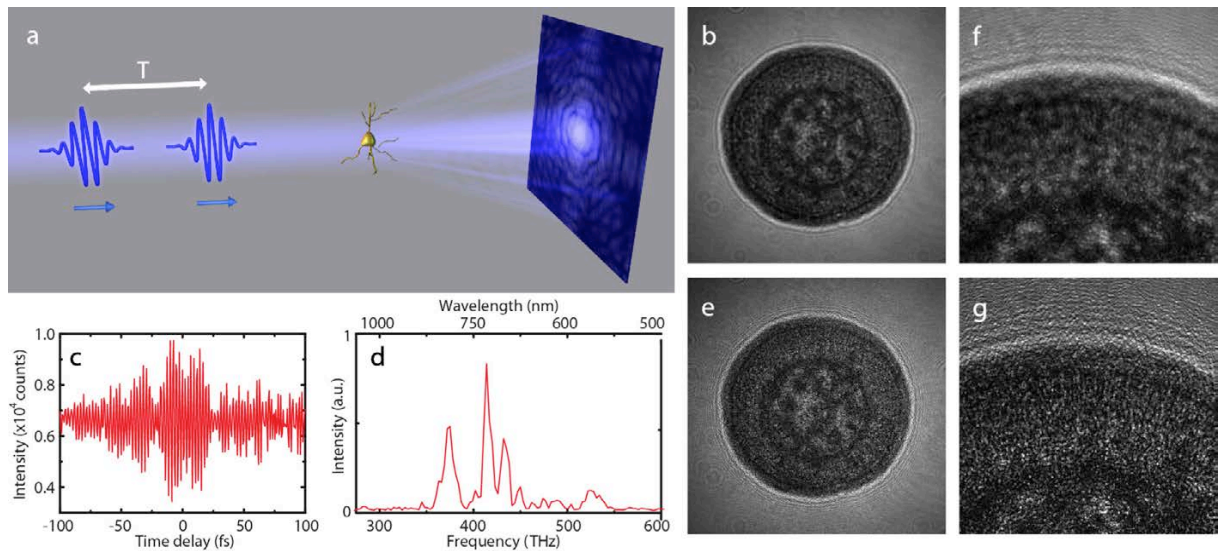

**Figure 1 | Principle of two-pulse Fourier-transform imaging. a**, A sample is illuminated with two coherent pulses, and a set of diffraction patterns is recorded as a function of time delay T in a lensless imaging geometry. **b**, Broadband Fresnel diffraction pattern of Convallaria majalis rhizome, recorded with an octave-wide visible light source. **c**, Typical signal recorded on a single pixel of 1b, as a function of time delay T . **d**, Fourier transform of the signal in 1c, showing the spectrum of the light that diffracted onto that specific pixel. By reconstructing such spectra for all pixels, spectrally resolved diffraction patterns can be extracted throughout the full source spectrum. **e**, Quasi-monochromatic Fresnel diffraction pattern of the same Convallaria sample as in 1b, obtained from a two-pulse imaging scan. Much finer diffraction features are visible. **f,g**, Enlarged images of the top part of 1b,e.

Here we demonstrate that by using two coherent time-delayed pulses, all bandwidth limitations on lensless imaging can be removed. No measurement of the source spectrum is required, and this spectrum may be ultra-broadband, highly structured and even exhibit significant intensity fluctuations (see SI). Our approach employs methods from Fourier-transform spectroscopy[14] to obtain spectrally resolved information from the recorded diffraction patterns themselves. We illuminate the sample with a time-delayed pulse pair (Fig. 1a), and record a series of diffraction patterns as a function of the time delay between the pulses.

For a first experimental realization of two-pulse imaging, we used a spatially coherent octave-spanning continuum as the light source. Two time-delayed pulses are generated in a scanning Michelson interferometer, and sent into a lensless imaging setup (see Suppl. Fig. 1 and Methods for details). We record two-pulse scans of Fresnel diffraction patterns of a fixed sample of Convallaria majalis (Lily of the Valley) rhizome, of which a broadband diffraction pattern is shown in Fig. 1b.

The key point in two-pulse imaging is that the signal from each individual camera pixel, recorded as a function of pulse-to-pulse time delay (Fig. 1c), encodes a Fourier-transform spectrum of the light scattered onto that specific pixel. Fourier-transforming the time-domain signal for all camera pixels then reveals the relative intensities of all spectral components scattered onto each pixel (Fig. 1d). From a single scan, spectrally resolved diffraction patterns can be reconstructed throughout the entire source spectrum, with a resolution limited only by the scanned time delay. A powerful feature of this approach, common to many Fourier-transform-based methods[14,15], is that the full source spectrum is used throughout the entire measurement, resulting in an efficient use of the available photon flux. Fig. 1e shows a narrow-band ($\Delta\lambda$ = 5 nm) diffraction pattern of the Convallaria sample at a wavelength of 695 nm, which has been extracted from the two-pulse imaging scan. Sharp diffraction features are clearly visible, which were washed out in the broadband images. Figs. 1f,g show expanded views of a part of the broadband and spectrally resolved diffraction patterns, respectively, to emphasize the improvement in diffraction fringe visibility.

Another important result that arises from our two-pulse method is that the spectral information can also be used for fast and robust Fresnel-domain phase retrieval. We have developed a novel iterative phase retrieval scheme, which explicitly uses the recorded multi-wavelength data to reconstruct the phase without the need for support constraints. In the Fresnel regime, wave propagation couples amplitude and phase of an electric field $E(x, y, z)$ through the Fresnel diffraction integral:

$$E(x,y,z) = \frac{e^{i2\pi z/\lambda}}{i\,\lambda\,z} \iint E(x',y',0)\, e^{\frac{i\pi}{\lambda z}[(x-x')^2 + (y-y')^2]} dx' dy' \qquad (1)$$

where propagation is along the $z$-coordinate, and $\lambda$ is the wavelength of the light. Fresnel domain iterative phase retrieval has been demonstrated using support-based methods[16,17] as well as measurements at multiple sample-to-camera distances[18,19]. Equation (1) states that Fresnel propagation depends on distance and wavelength in an identical way (aside from a global phase factor), allowing us to exploit our spectrally resolved diffraction data to 'propagate' between different spectral components, as schematically depicted in Fig. 2a. This scheme does not require sample or camera movement, and only relies on measured data rather than specific sample assumptions or support constraints. It converges reliably and works for extended samples, for which support-constraint-based algorithms fail. While other methods such as keyhole imaging[20] and ptychography[21] have been used successfully to image parts of extended samples, our multi-wavelength approach does have the advantage that spatially confined illumination is not required.

A demonstration of multi-wavelength phase retrieval is presented in Fig. 2, where we reconstruct an image of the Convallaria sample based on diffraction images as shown in Fig. 1e. The multi-wavelength phase retrieval algorithm (Fig. 2a) results in a high-quality image reconstruction, which is displayed in Fig. 2b. While the sample fills most of the field-of-view (FOV), our multi-wavelength algorithm enables image reconstruction at instrument-limited resolution (see Methods), clearly showing the individual cells as well as some sub-cellular features (Fig. 2c). After the phase has been retrieved, all spectrally resolved images can be propagated to the object plane and averaged to improve signal-to-noise, resulting in an efficient use of the full source spectrum. The resolution of such a reconstruction is then determined by the imaging geometry and the weighted mean of the spectral bandwidth. The main prerequisite of our multi-wavelength phase retrieval approach is that the phase profile of the object is identical for the spectral components used in the reconstruction. However, if this assumption is not met, support-based iterative phase retrieval schemes can still provide image reconstruction[11,17], which would enable spectroscopic imaging.

In addition to transmission imaging, lensless imaging in a reflection geometry[7] is important for many applications in nanoscience and technology[22,23]. The need for a finite support is particularly challenging to satisfy for reflection geometries[7], and forms a barrier for practical lensless reflection imaging. In contrast, our multi-wavelength approach enables imaging of large areas of extended samples in a reflection geometry. Fig. 2d shows a spectrally resolved Fresnel diffraction pattern, retrieved from a two-pulse scan in reflection of a USAF 1951 test target. The reconstructed image obtained through multi-wavelength phase retrieval is shown in Fig. 2e, demonstrating the ability to image large reflective samples without support requirements.

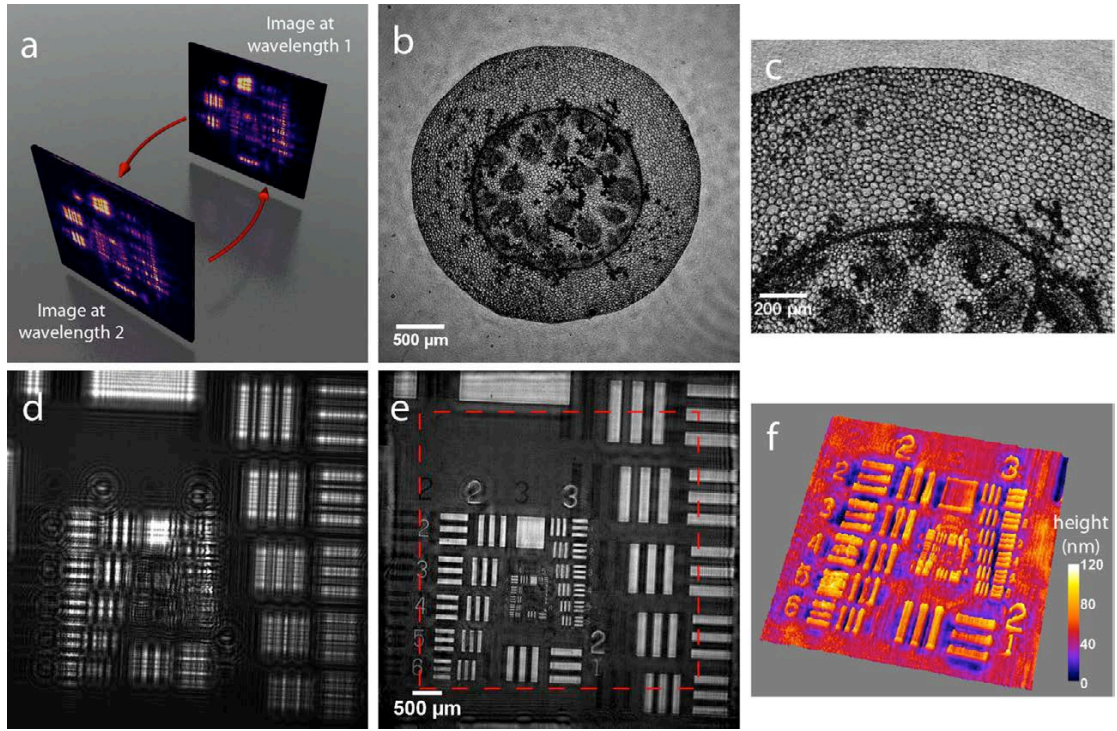

**Figure 2 | Multi-wavelength phase retrieval in the Fresnel regime. a**, Graphic representation of the multi-wavelength phase retrieval algorithm. Iterative propagation is performed between Fresnel diffraction patterns recorded at the same distance, but with different wavelengths. **b**, Resulting image after 25 iterations of the multi-wavelength iterative phase retrieval, using diffraction patterns at 5 different wavelengths. **c**, Zoom-in on a part of the reconstructed image, clearly showing the individual cells. **d**, Quasi-monochromatic Fresnel diffraction pattern of a 1951 USAF test target, recorded in a reflection geometry and obtained through a two-pulse Fourier-transform scan. **e**, Resulting image after 25 iterations of the multi-wavelength iterative phase retrieval, using diffraction patterns at 5 different wavelengths. Some negative (dark) features are also observed, which are due to reflections from the sample back surface. The dashed red line marks the area in which all wavelengths contribute to the reconstruction. **f**, Reconstructed height map from part of the USAF target surface from 2e, showing the chrome structures on the glass substrate.

As the field-of-view in lensless imaging is wavelength-dependent, not all spectrally resolved images contribute constructively to the reconstruction near the edges. This leads to slightly deformed edges around the image, as well as a 'wrapping' artefact where bright structures at the edge of the FOV appear on the opposite edge of the reconstructed image. The FOV in which all spectral components contribute to the reconstruction is determined by the shortest wavelength, and is shown as the dashed red line in Fig. 2e. In general, the iterative phase retrieval algorithm provides accurate phase information, which can be used (assuming that the signal is purely a surface reflection) to reconstruct a height map with ~20 nm resolution of the patterned structures on the sample surface (Fig. 2f).

To demonstrate that our two-pulse imaging approach extends to shorter wavelengths, we perform an imaging experiment with a HHG source (see Methods). The pulse pair is produced in the near infrared with a split-wavefront interferometer before the HHG setup, using quartz wedges to scan the time delay (Suppl. Fig. 2). The interferometer is aligned such that the two pulses propagate collinearly but are spatially separated at the focus. This geometry ensures that ionization due to the first pulse does not influence the HHG process for the second (time-delayed) pulse, allowing linear Fourier-transform scans at HHG frequencies[24]. The HHG pulses overlap spatially at the imaging target due to their beam divergence (see SI text for details).

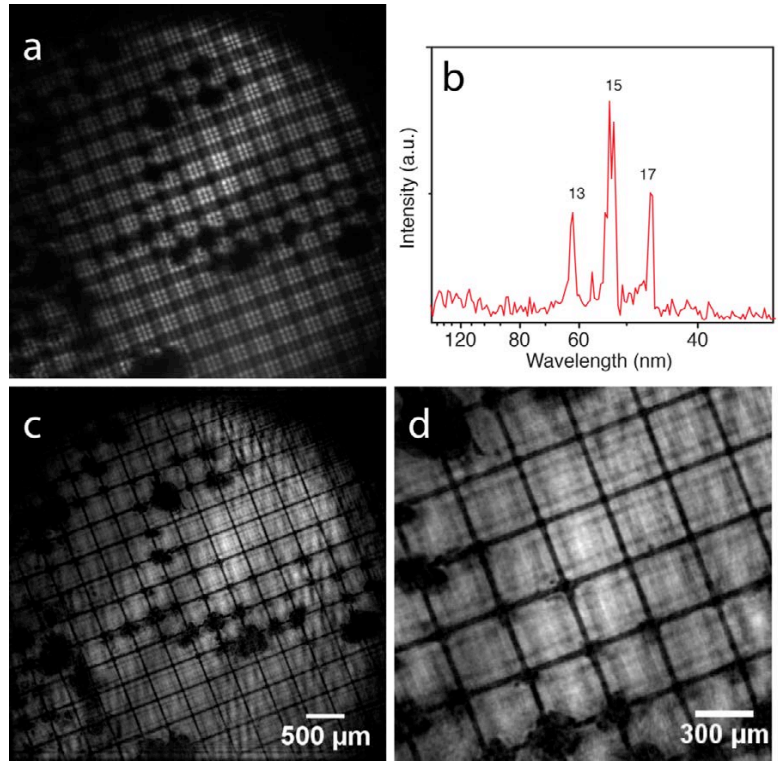

**Figure 3 | Broadband lensless imaging with extreme-ultraviolet radiation. a**, Broadband diffraction pattern of a Nickel grid on a 300 nm thick Al foil, recorded using harmonics 13, 15 and 17. **b**, Spectrum extracted from the two-pulse Fourier-transform scan. **c**, Retrieved image of the grid after 10 iterations of the multi-wavelength phase retrieval algorithm, using diffraction patterns at 3 different harmonics as input. The regular features of the grid are clearly visible, along with several damage spots induced by intense laser irradiation in earlier experiments. **d**, Enlarged view of part of the image in c, showing the 20 μm wide grid lines.

We use the full HHG flux directly to image a Nickel grid, with 20 μm wide bars, attached to a 300 nm thick Aluminium foil. Figure 3a displays a broadband diffraction pattern of the Nickel grid recorded with the transmitted HHG spectrum. Figure 3b shows the HHG spectrum used in these imaging experiments, as derived from the two-pulse scan itself. Three harmonics are present, spanning a wavelength range between 47 and 63 nm. Lower harmonics are not transmitted by the Al foil and the Xenon gas, while the HHG phase-matching cutoff is near harmonic 17. Spectrally resolved diffraction patterns at harmonics 13, 15 and 17 are used as input for the multi-wavelength phase retrieval algorithm. Due to the transverse displacement between the two HHG pulses, the recorded data contain two displaced copies of the object's diffraction pattern. This complication can be accounted for by incorporating an additional deconvolution step in the image reconstruction procedure (see SI text for details). Small angular deviations between the two beams can also be handled, enabling two-pulse imaging even with non-collinear beams from a split-wavefront interferometer[25,26].

The resulting reconstructed image of the Ni grid, with a diffraction-limited resolution of 6.7 μm, is shown in Figs. 3c and 3d. In addition to the regular Ni grid, several dark damage spots are observed, which were caused in earlier experiments with intense laser pulses that partially melted the Nickel structures. While the diffraction limit in the current proof-of-concept experiment was relatively low due to geometrical constraints, a much higher resolution can readily be achieved in Fresnel diffractive imaging by illuminating the sample with a curved wavefront[16,20], which is fully compatible with our two-pulse imaging approach.

We find that two-pulse imaging enables robust and accurate lensless imaging with broadband and unstable spectra, without a priori knowledge of the spectrum for image reconstruction. This is a situation often encountered with HHG sources, where efficient use of the available photon flux is essential for practical imaging applications. We have performed numerical simulations to study the influence of intensity and timing variations in more detail (SI text), which indicate that good-quality images can be obtained even in the presence of significant noise. From our measurements, we find a path length stability of 8 nm between the pulses (SI text), already allowing imaging at wavelengths down to 16 nm. With additional stabilization measures, an extension to the soft-X-ray domain is well feasible. We therefore foresee that two-pulse imaging will find widespread application in the development of compact table-top soft-X-ray microscopes[27–30], thus providing new possibilities for e.g. structural biology and nanotechnology by enabling label-free ultrahigh-resolution microscopy in a laboratory-scale environment.

**Methods**

**Visible light two-pulse imaging** Coherent white-light continuum pulses are produced by launching ultrashort laser pulses into a photonic crystal fiber. A time-delayed pulse pair is produced with a Michelson interferometer, with one of the end mirrors mounted on a closed-loop piezo stage. The pulse pair is then used to illuminate a sample, and the diffracted light is recorded with a 14-bit CCD camera (Suppl. Fig. S1). We typically acquire 500 individual images with 0.01 to 1 millisecond exposure time, while increasing the time delay in 0.67 fs (200 nm) steps. We recorded Fresnel diffraction patterns of the Convallaria sample with the CCD at a distance of 19 mm behind the sample, which corresponds to a diffraction limit of ~2 μm. The CCD pixel size of 4.54 μm was the limiting factor in these experiments, giving rise to a 4 μm pixel step size in the final reconstructed image.

**HHG two-pulse imaging** We use pairs of 0.2 mJ, 40 fs pulses emitted by a Ti:Sapphire amplifier system, focused into a semi-infinite gas cell containing Xenon for HHG. We used a split-wavefront interferometer to produce the two pulses at the fundamental wavelength, which were used for HHG at spatially separated locations inside the gas cell. Scanning the time delay was performed by moving a wedge into one half of the beam, enabling scans with sub-nm position accuracy (SI text). We recorded images as a function of time delay in 512 steps of 44 attoseconds (13.2 nm path length change per step).

**Spectrally resolved image retrieval** Spectrally resolved images are obtained from a two-pulse scan by loading the full (x,y,t)-dataset into a 3D array and performing a 1D-FFT as a function of time delay for each individual camera (x,y)-pixel. This directly yields a (x,y,f)-dataset, where each spatial 2D image is now a spectrally resolved diffraction pattern at frequency $f = c/\lambda$, with a spectral resolution of $1/T$. Each of these diffraction patterns can then be further processed with the algorithms available for lensless image reconstruction. The spectral resolution $\Delta\lambda$ of the individual reconstructed images is only limited by the total scanned time delay T, and can be expressed as $\Delta\lambda/\lambda = \lambda/(c\,T)$, where c is the speed of light and $\lambda$ is the wavelength of the light. Similar to Fourier-transform spectroscopy, the maximum step size between images is limited to $\lambda/2$ by the Nyquist sampling criterion.

**Multi-wavelength phase retrieval** For our multi-wavelength Fresnel reconstruction method, phase retrieval is performed in a Gerchberg-Saxton type iterative scheme, where the intensity data from the first wavelength is propagated to the next wavelength through evaluation of the Fresnel propagation equation (1). After propagation, the phase information is retained, while the intensity is replaced with the measured intensity at this new wavelength. We typically select 2-5 images separated by 5-20 THz in frequency from the spectrally resolved dataset for iterative phase retrieval. The algorithm converges rapidly, typically requiring only 10-50 iterations to reach a final solution.

**Acknowledgements** The authors thank Prof. P.S. Carney for insightful discussions, and Prof. M.L. Groot for providing access to her Ti:Sapphire amplifier system. This work is (partly) financed by an NWO-groot investment grant of the Netherlands Organisation for Scientific Research (NWO) and Laserlab Europe (JRA Bioptichal). S.W. acknowledges support from NWO Veni grant 680-47-402.

**Author contributions** S.W. devised the concept of two-pulse Fourier-transform lensless imaging. All authors took part in the design of the experiments. S.W. V.T.T and D.W.E.N. built the setups and performed the experiments. S.W. and V.T.T. performed the data analysis. K.S.E.E. supervised the project. All authors contributed to interpretation of the results and writing of the manuscript.

**Additional information** The authors declare no competing financial interests. Correspondence and requests for materials should be addressed to S.W. (s.m.witte@vu.nl) or K.S.E.E. (k.s.e.eikema@vu.nl).

# Supplementary information

**SETUP FOR VISIBLE LIGHT EXPERIMENTS**

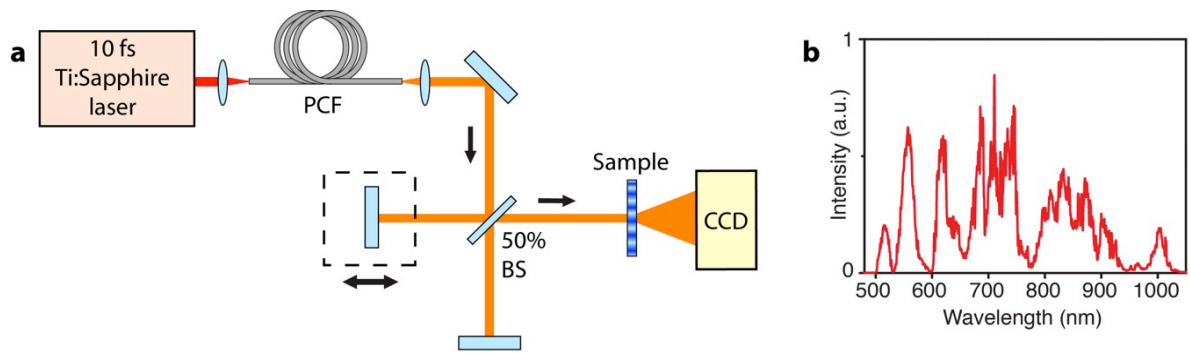

**Supplementary Figure 1: Setup for lensless two-pulse imaging. a,** White-light continuum pulses are produced by launching 10 fs, 2.5 nJ pulses from a modelocked Ti:Sapphire laser, running at 80 MHz repetition rate, into a photonic crystal fibre (PCF) with a 2.3 μm core diameter. Supercontinuum generation in the fibre produces a highly structured white light continuum that spans an optical octave, as shown in **b**. A pulse pair with variable time delay is produced by a Michelson interferometer, with one of the end mirrors mounted on a closed-loop piezo-driven translation stage.

**SETUP FOR HHG EXPERIMENTS**

A schematic of the setup used for two-pulse lensless imaging with HHG radiation is given in Suppl. Fig. 2. We start with intense ~40 fs pulses emitted by a Ti:Sapphire amplifier system (Spectra Physics Spitfire Ace). From this beam, a pulse pair with variable time delay is produced by passing half the incident beam through a pair of wedges. The time delay T can be scanned by moving one of the wedges further into the beam. This particular interferometer implementation is highly stable: it is nearly common path, and there are no mirrors in the setup that only reflect one of the pulses. These properties eliminate most of the noise caused by vibrations, air flow and acoustics present in conventional Michelson interferometers. We used a 1° wedge for scanning the delay, mounted on a closed-loop piezo-stage with 5 nm resolution and 500 μm scan range. A 100 nm movement of the stage resulted in a 1.32 nm optical path length difference between the pulses, enabling scans with sub-nm step size.

A single wedge in the other interferometer arm introduces an angle between the beams. The lens (focal length F), that focuses the pulses into the HHG gas cell, is placed at a distance F behind the point where the beams cross. This geometry ensures that the two beams run parallel to each other behind the lens. Both beams focus near the end of the gas cell with a small transverse displacement between the focal spots. In a geometry where the focal spots overlap, the ionization induced by the first pulse would distort the phase-matching conditions for the second pulse, resulting in time-dependent intensity modulations and nonlinear phase shifts in this second pulse. This would severely distort a Fourier-transform scan. Because of the transverse separation that we introduce in our setup, both pulses will produce HHG independently, allowing clean Fourier-transform images to be recorded[1]. The separation between the spots is determined by the angle of the wedge and the focal length F, and is set at 0.2 mm. The focal length F is 350 mm, and the focal spot diameter is 40 μm

FWHM. Harmonics are produced in Xenon gas at ~50 mbar pressure in the gas cell, with an energy of 0.2 mJ in each pulse. Behind the cell, a differential pumping stage is used to prevent re-absorption of the produced HHG.

We use the full HHG flux directly to image a Nickel grid attached to a 300 nm thick Aluminium foil. The Al foil reflects the fundamental beam, while transmitting radiation below 70 nm wavelength (harmonics 13 and higher). In this experiment, the Nickel grid is placed at 40 cm behind the focus. The diffracted light is detected in a transmission geometry using an XUV-sensitive CCD camera (Andor Technology) with 1024x1024 pixels, a pixel size of 13 μm, and a bit depth of 12 bits. In a typical Fourier-transform scan, 512 diffraction images are recorded as a function of time delay. Between consecutive images, the time interval is increased in steps of 44 attoseconds (step size 13.2 nm). At each time step, 5 frames are recorded with 0.3 second exposure time and averaged to improve the signal-to-noise ratio.

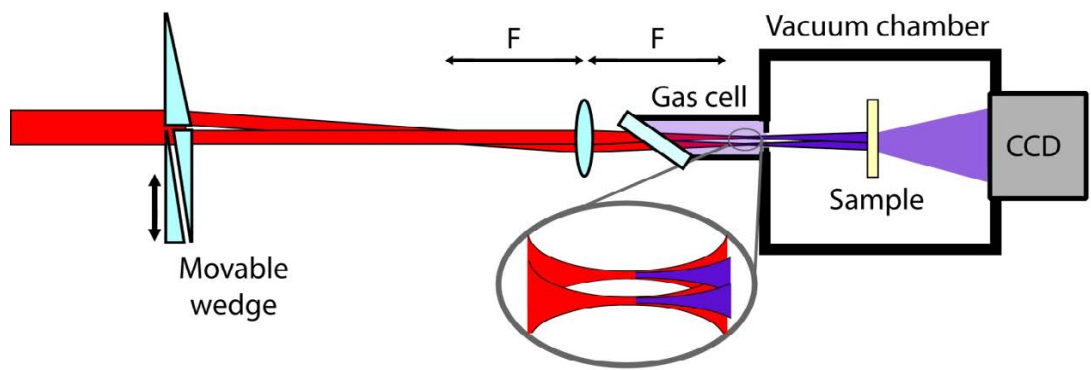

**Supplementary Figure 2: Setup for two-pulse Fourier-transform lensless imaging with high-harmonic radiation.** A variable time delay is introduced by inserting a wedge into the beam (note that this drawing is only for schematic purposes: the actual orientation of the wedges is rotated by 90 degrees with respect to this drawing). See SI Text for details.

## INTERFEROMETER STABILITY MEASUREMENTS

To assess the stability of our interferometer, we use the data from our two pulse scan itself. We record 5 images consecutively at each time delay, and we compare these images to obtain an estimate of the path length fluctuations between the two pulses. Due to the noncollinear geometry of our experiment, the individual images contain spatial interference fringes. We isolate this interference pattern by a 2D spatial Fourier transformation, filtering the spatial frequency corresponding to this interference pattern, and inverse transformation. We can then extract the phase of the interference at a single pixel. An estimate for the path length jitter is obtained by calculating the average phase of each group of 5 measurements at a fixed position, and then calculating the difference of the phase of each individual measurement with this average. Since a $2\pi$ phase shift corresponds to a path length change of one average wavelength, this phase shift is a direct measure of the path length fluctuations.

Performing such an analysis on an entire two-pulse imaging dataset provides a histogram of the path length fluctuations during the scan, of which an example is given in Suppl. Fig. 3. From this data, we find that the path length fluctuations in our scan follow a Gaussian distribution with a FWHM width of 8 nm. We have repeated this analysis on multiple datasets, yielding identical results. From this analysis and the results of our numerical simulations (see SI text below) we conclude that imaging with wavelengths down to 16 nm should already be possible with the level of stability of our current setup. With further stabilization measures, such as active feedback and placing the interferometer in vacuum, these fluctuations can be improved to the nm level, enabling two-pulse imaging in the water-window spectral range.

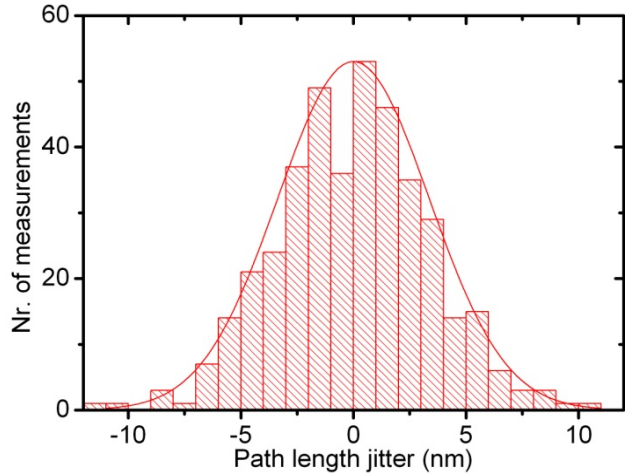

**Supplementary Figure 3: Measurement of the path length fluctuations during a scan.** Histogram of the phase deviations observed during a two-pulse imaging experiment. The width of the Gaussian distribution is 7.9 nm FWHM. See SI text for details.

**MULTI-WAVELENGTH FRESNEL IMAGE RECONSTRUCTION PROCEDURE**

For our multi-wavelength Fresnel reconstruction method, phase retrieval is performed in a Gerchberg-Saxton type iterative scheme[2], where the intensity data from the first wavelength is propagated to the next wavelength using the Fresnel propagation equation (1). After propagation, the phase information is retained, while the intensity is replaced by the measured intensity at this new wavelength.

For the visible light experiments, we typically select 2-5 images separated by 5-20 THz in frequency from the full spectrally resolved dataset for iterative phase retrieval. The algorithm converges rapidly, typically requiring only 10-50 iterations to reach a final solution. For all samples that we imaged, the multi-wavelength phase retrieval algorithm always found the correct solution, without stagnation in local minima.

For the HHG experiments, the transverse displacement between the pulses requires an additional deconvolution step. In this geometry, the pulses produce two identical diffraction patterns, which are transversely displaced. As our spectrally resolved diffraction images are obtained from a Fourier-transform, which gives the cycle-averaged spectral intensity at each pixel, no interference between these displaced diffraction patterns is observed. The image at each wavelength can therefore be interpreted as the convolution between a single diffraction pattern of the object and a pair of Dirac delta functions. The distance between these Dirac delta functions is calculated from the imaging geometry and the initial separation between the focal spots in the gas cell. The single diffraction patterns can then be retrieved from the spectrally resolved images by deconvolution. After this deconvolution step, the resulting diffraction patterns can be used in the multi-wavelength phase retrieval algorithm. Note that this deconvolution step also works even if the two pulses do not propagate collinearly. Therefore, even interferometers that introduce a finite angle between the pulses, such as split-wavefront devices[3,4], can be used for our two-pulse lensless imaging approach. The main requirement in such a geometry is that interference remains visible for the main part of the diffraction pattern, i.e. diffracted light from both beams should ideally be present at most of the CCD pixels. This is usually the case for Fresnel diffraction, but not necessarily so for far-field (Fraunhofer) diffraction imaging.

For the HHG phase retrieval we use images at harmonics 13, 15 and 17 (center wavelengths 62 nm, 53 nm and 47 nm, respectively). At each harmonic, we filter out a spectral bandwidth of 1.5 nm, which suffices to achieve diffraction-limited resolution in this experiment. Typically, 10 iterations of the algorithm are required for stable convergence.

**TWO-PULSE IMAGING IN THE PRESENCE OF NOISE: SIMULATIONS**

In Fourier-transform spectroscopy, the signal at any frequency component is encoded as a sinusoidal signal with a constant oscillation period in the time-delay scan. Therefore, only temporal intensity fluctuations with a frequency that matches this signal oscillation will add noise to the image at this particular frequency component.

To investigate the amount of temporal intensity fluctuations that can be tolerated, we performed numerical simulations of our two-pulse scans at realistic experimental parameters, but with various types of noise added to either one or both pulses. We then performed our multi-wavelength phase-retrieval procedure, to assess the influence of specific types of noise on the quality of the final reconstructed image. Supplementary Fig. 4 shows the results of these simulations. We used a typical image of part of a resolution test target as the input object, and calculated diffraction patterns at 0.25 metre propagation distance, with an octave-wide input spectrum between 2.25 and 4.5 PHz (67 – 133 nm wavelength range). The red box in Suppl. Fig. 4 shows the input spectrum, input object, and a typical calculated diffraction pattern at 133 nm wavelength.

We simulated two-pulse scans for several cases, being 1) perfectly stable pulses, 2) certain amounts of timing jitter between the pulses, 3) spectral intensity fluctuations in both pulses simultaneously, and 4) spectral intensity fluctuations in only one of the pulses. In each case, 512 steps with 20 nm step size (66.7 attoseconds) are taken, symmetrically around zero time delay. For each situation, we plot the resulting temporal interference at a single pixel (from the top bright part of the '4' in the diffraction pattern), the retrieved spectrum averaged over the full spectrally resolved dataset after the Fourier transformation step, and the final image of the object retrieved by multi-wavelength phase retrieval (using 3 diffraction patterns at wavelengths of 126.7 nm, 92.1 nm and 69.9 nm as input data).

For the case of stable pulses, the time delay scan shows a clean interference pattern, the source spectrum is retrieved with good quality (aside from smoothed edges due to the finite spectral resolution), and the reconstructed image of the object is of high quality.

Next, we simulated a scan in which introduced a random timing error to each scan step, taken from a Gaussian distribution with a FWHM of 50 attoseconds (15 nm path length). Such a timing jitter corresponds to nearly a quarter cycle of the shortest wavelength in the source spectrum. While the time delay scan still looks good, a significant amount of white noise is introduced onto the retrieved spectrum. Nevertheless, the shape of the spectrum is retrieved correctly, and the reconstructed image seems to suffer only a minor contrast decrease. Increasing this timing jitter to 100 attoseconds (30 nm path length) results in more severe errors. The noise level on the spectrum has increased to the level of the signal. As a result, the reconstructed image also shows a significant background noise, although a faithful reconstruction of the object is still obtained. This is quite remarkable, since this timing jitter corresponds to nearly half a cycle of the shortest wavelength in the spectrum.

To simulate spectral intensity jitter, we divided the spectrum into 50 components. At each scan step, we gave each individual component a random intensity between 50 and 100%. This would simulate a situation where a single fluctuating pulse is split in two for the measurement. The reconstructed spectrum is as expected, and the retrieved image of the object is of high quality. To study the situation where two pulses experience independent fluctuations (as could be the case for HHG with separated generation zones), we simulated a scan with such spectral intensity fluctuations in only one of the pulses. Again in this case, both the retrieved spectrum and reconstructed image are good representations of the input parameters. It should be stressed that the simulated intensity fluctuations are excessive even for HHG sources, and can easily be minimized by averaging over multiple laser pulses at each scan step. The simulated timing variations are also significantly larger than experimentally observed. The exact influence of the noise on the contrast and resolution of the retrieved image depends on the object under investigation, and is therefore difficult to quantify. Yet the two-pulse imaging approach is found to be remarkably robust, and results in a faithful reconstruction of the object even in the presence of significant noise sources. This gives confidence that the method can be scaled to significantly shorter wavelengths than presented in the current work.

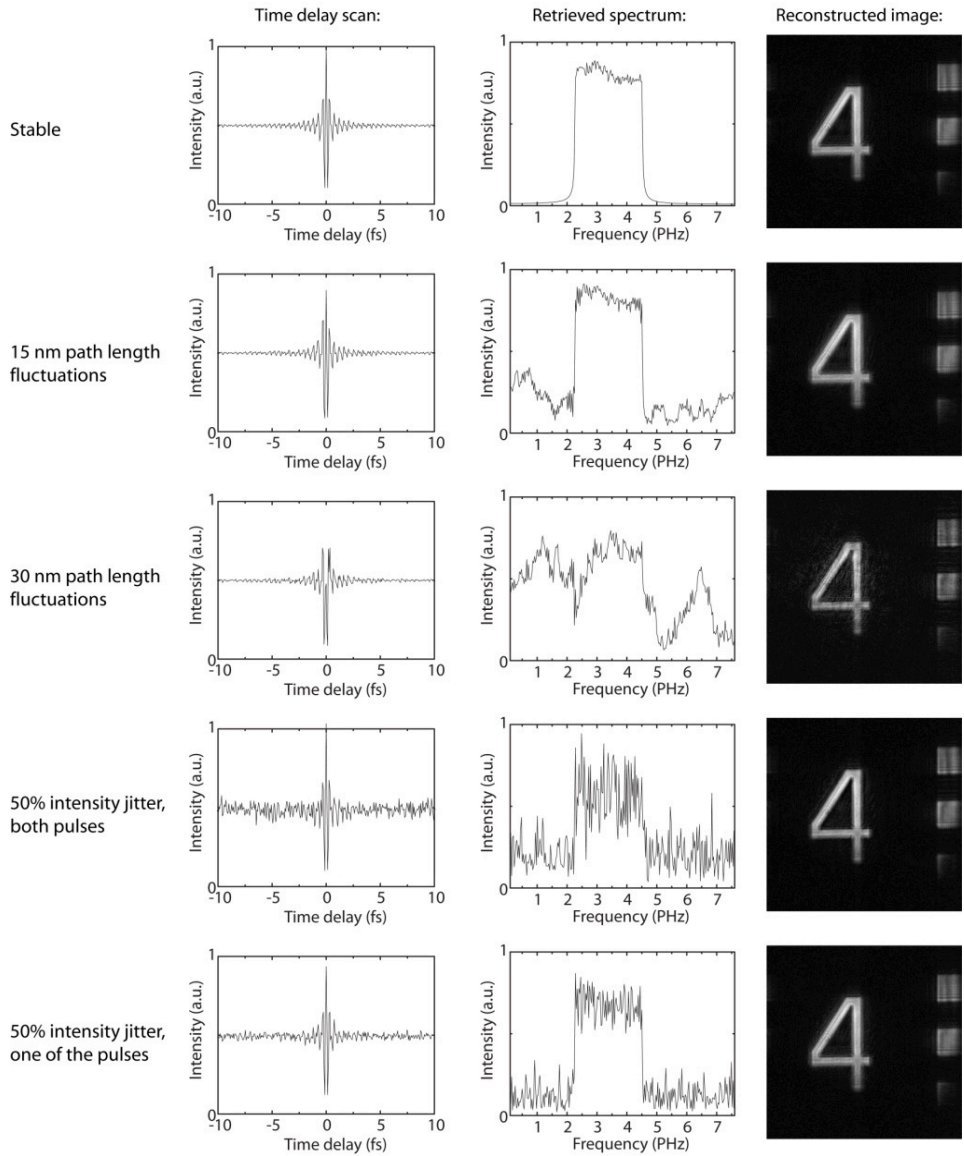

**Supplementary Figure 4: Effects of different types of noise on the two-pulse scan and the resulting image reconstruction.** The input spectrum and image object are displayed in the red box at the top, along with a simulated diffraction pattern at one wavelength. In the respective rows, five different simulation results are shown. For each simulation, the left column displays the intensity at a single pixel as a function of pulse-to-pulse time delay, the middle column displays the spectral intensity after Fourier transformation, and the right column shows the reconstructed image from the multi-wavelength phase retrieval algorithm. See SI text for details.

14